\documentstyle[11pt]{article}
\begin{document}
\begin{center} 
{\bf A Speckle Experiment during the Partial Eclipse}
\end{center} 
\vspace{0.4cm}

\begin{center} 
{\bf S. K. Saha, B. S. Nagabhushana, A. V. Ananth and P. Venkatakrishnan} \\
Indian Institute of Astrophysics, Bangalore 560034 \\
\end{center} 
\vspace{0.4cm}

\noindent
{\bf Abstract} 
\vspace{0.3cm}

An experiment for the speckle reconstruction of solar features was developed 
for observing the partial eclipse of the sun as viewed from Bangalore on 
October 24, 1995. No data could be obtained because of cloudy sky but the 
experimental details are described.
\vspace{0.3cm}

\vspace{0.3cm}

\noindent
{\bf 1. Introduction} 
\vspace{0.3cm}

Many problems in solar physics require information about the solar surface
features at the highest possible angular resolution. The earth's atmosphere
blurs the images. In the case of night-time astronomy, image reconstruction
techniques have been developed (Labeyrie, 1970, Knox and Thompson, 1974,
Weigelt, 1978), that take advantage of a nearby point source as a reference
for the reconstructions. This is not possible for the solar features. The lunar
limb provides a sharp edge as a reference object during solar eclipses. A
solar speckle experiment was planned for observing the solar eclipse of Oct. 24,
1995 visible as a partial eclipse from Bangalore.
\vspace{0.3cm}

\noindent
{\bf The Instrumentation}
\vspace{0.3cm}

A Carl-Zeiss 15 cm Cassegrain-Schmidt reflector was used as the telescope for
the experiment. To prevent heating of the optics, an aluminised glass plate was 
fixed in front of the telescope that reflected back 80 percent of the sunlight 
and transmitted only 20 percent. A 3~nm passband filter centered at 600~nm was
placed after this, followed by another polaroid mounted on a rotatable holder
as shown in Figure 1. The amount of light falling on to the camera can be 
adjusted by rotating the second polaroid. A pin-hole of 1~mm diameter was 
placed at the focal plane for isolating a small field-of-view. A microscope
objective reimages the pin-hole on to the camera. The camera is a EEV CCD 
camera operated in the TV mode. The images can be acquired with exposure time
of 20~ms using a Data Translation$^{TM}$ frame-grabber card DT-2861 and
subsequently stored on to the hard disk of a PC/AT computer.
\vspace{0.3cm}

\noindent
{\bf Result}
\vspace{0.3cm}

No images of the partially eclipsed sun could be acquired due to unfavourable
weather conditions at Bangalore on Oct. 24, 1995. The image reconstruction
involves the treatment of both amplitude errors and phase errors. The 20~ms
exposure time is small enough to preserve phase errors. Any of the schemes
for phase reconstruction that satisfactorily reproduces the lunar limb would
be valid for solar features close to the limb, i. e., within the isoplanatic
patch. Also, the limb reconstruction would be valid only for phase distortions
along one dimension (in a direction normal to the lunar limb). In spite of
these shortcomings, the limb data would have provided additional constraints
for techniques like blind iterative deconvolution.
\vspace{0.3cm}

\noindent
{\bf Acknowledgments} 
\vspace{0.3cm}

The personnel of Bangalore workshop, in particular Messrs T. Periyanayagam and
N. Thimmaiah, provided excellent support for fabricating the instrument. Mr.
V. Gopinath of photonics division, Mr. R. M. Paulraj of mechanical design
section, Mr. K. Padmanabhan of electrical division, A. S. Babu of Electronics
division also helped in various ways. The enthusiastic help rendered by 
Messers V. Krishnakumar, K. Sankarasubramanian and R. Sreedharan (all Ph. D
students) during the testing phase is also acknowledged.
\vspace{0.3cm}

\noindent
{\bf References} 
\vspace{0.3cm}

\noindent
Knox K.T. \& Thompson B.J., 1974, Astrophys. J. {\bf 193}, L45. \\ 
\noindent
Labeyrie A., 1970, Astron. Astrophys., {\bf 6}, 85. \\
\noindent
Weigelt G., 1978, Appl. Opt. {\bf 17}, 2660.
\vspace{0.3cm}

\noindent
{\bf Figure Caption}
\vspace{0.3cm}

\noindent
{\bf Figure 1}: Schematic layout of the instrument. 
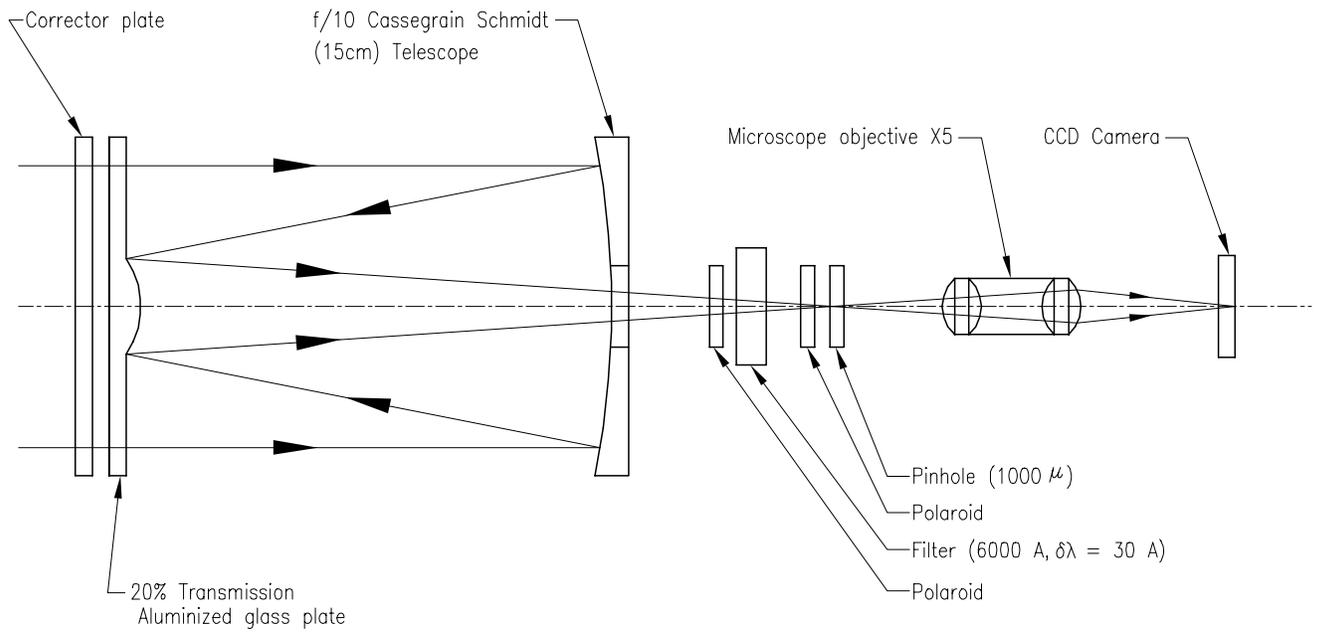
\end{document}